\documentclass[11pt]{article}

\usepackage{arxiv}

\usepackage{cite}
\usepackage{amsmath,amssymb,amsfonts}
\usepackage{algorithmic}
\usepackage{graphicx}
\usepackage{textcomp}
\usepackage{xcolor}
\usepackage{hyperref}
\usepackage{tabularx}
\usepackage{multirow}

\title{Energy-Efficient CNN Acceleration with MSDF Digit-Serial Arithmetic on FPGA}

\author{
  Muhammad Usman,
  Yousef Sadegheih,
  Dorit Merhof \\
  Faculty of Informatics and Data Science, University of Regensburg, 93053 Regensburg, Germany \\
  \texttt{\{muhammad.usman,yousef.sadegheih,dorit.merhof\}@ur.de}
}
\begin{document}
\maketitle

\begin{abstract}
This paper presents an energy-efficient hardware acceleration of the convolutional layers in the U-Net architecture for image segmentation, implemented on FPGA. While digit-serial arithmetic, particularly most-significant-digit-first (MSDF) techniques, offers a compact hardware footprint, it suffers from initial latency before producing the first output digit. This delay accumulates in cascaded operations like multiplication followed by addition, where each unit introduces its own startup overhead. To overcome this, we propose a merged multiply-add (MMA) architecture that fuses these operations into a unified pipeline. Instead of incurring separate delays, the MMA introduces a single streamlined latency per iteration, shorter than the combined latency of conventional cascaded units, resulting in enhanced throughput and efficiency. The MMA units are designed to process spatial input depths in parallel, achieving significantly higher performance than both standalone MSDF-based and conventional designs. We evaluate the proposed design using U-Net as a target application. Despite operating at a lower frequency than a CPU, the FPGA-based accelerator achieves up to an order of magnitude higher energy efficiency, delivering up to $15.14$ GOPS/W compared to $1.93$ GOPS/W for CPU-based inference. The design also shows approximately $9\times$ reduction in energy consumption compared to MSDF-based FPGA implementations. These results highlight the efficacy of the merged arithmetic approach for resource-constrained, latency-sensitive edge applications in medical imaging and computer vision.
\end{abstract}


\section{Introduction}\label{sec:Intro}
Accurate brain tumor segmentation plays a vital role in medical imaging by assisting clinicians in diagnosis, treatment planning, and postoperative assessment. U-Net, a convolutional neural network (CNN)-based architecture, has become the de facto standard in biomedical image segmentation due to its ability to capture both contextual and localization features, making it well-suited for delineating complex anatomical structures such as gliomas, meningiomas, and metastases in brain MRI scans \cite{azad2024medical}. The computation in U-Net is primarily dominated by convolution operations, which demand high computational resources and energy when implemented on CPU and GPU \cite{cheng2022efficient}, and poses challenges for embedded deployment.

FPGAs (Field-Programmable Gate Arrays) have emerged as a promising hardware platform for accelerating CNN workloads \cite{liu2024review}. They offer inherent advantages such as parallelism, low power consumption, and the ability to tailor architectures to specific applications. These characteristics make FPGAs well-suited for edge AI applications and embedded systems where energy efficiency and adaptability are critical, such as real-time diagnostic systems \cite{zhang2024appq}.

Nearly $85\%$ of the computation time in CNN-based models is spent on inner product calculations. To accelerate these operations, various inner product computation units have been developed with a focus on optimizing power, performance, and silicon area. Among these, bit-serial computation techniques have shown significant promise. By reducing interconnect bandwidth and simplifying hardware complexity, bit-serial methods are especially well-suited for low-power and area-constrained systems \cite{chhajed2022bitmac}. Unlike traditional fixed-width multiplication, bit-serial architectures perform multiplication sequentially, one bit at a time, accumulating results into partial sums. This breakdown simplifies hardware design, reduces logic complexity, and allows dynamic precision control, which is especially useful for applications requiring flexible accuracy and power trade-offs. Moreover, bit-serial processing can lead to substantial energy savings by eliminating redundant computation and minimizing data movement \cite{li2022bitcluster}. To this end, we propose a digit-serial MSDF arithmetic-based merged multiply-add (MMA) unit to accelerate convolution operations in U-Net, offering enhanced throughput with low hardware overhead. The key contributions of this work are as follows:
\begin{itemize}
    \item A novel MSDF-based multiply-add architecture fuses digit-serial multiplication and accumulation into a single pipeline. Unlike conventional designs with separate multiplier and adder delays, the proposed unit incurs only a single combined delay. This reduces per-iteration latency and improves throughput.
    \item The MMA units are structured to process multiple spatial input depths in parallel, significantly increasing the throughput of convolution operations central to U-Net.
    \item An energy-efficient FPGA-based accelerator is implemented for the convolutional layers in U-Net, demonstrating its applicability to resource-constrained, latency-sensitive edge applications.
    \item The performance of the proposed design is compared with CPU, bit-parallel, conventional and MSDF arithmetic-based convolution computation units in terms of latency, throughput, area and energy efficiency.
\end{itemize}

\section{MSDF Arithmetic}\label{sec:MSDF}
Most-significant digit first (MSDF) arithmetic has emerged as a promising technique in CNN acceleration \cite{ibrahim2025usefuse}. It enables digit-wise computation that supports early termination and natural pipelining, which are particularly advantageous for reducing latency and improving throughput in deep CNN inference. The computation of most significant digit first offers multiple benefits, including early termination, low latency, and progressive refinement of results. It naturally supports approximate computing, as early digits often provide enough precision for decision-making \cite{usman2023low}. Furthermore, its digit-wise pipelining structure enables overlapping computation and communication, which helps maximize throughput. Unlike traditional multiply-accumulate (MAC) units, MSDF-based inner product units are lightweight and can be deeply pipelined, making them ideal for FPGA fabric without reliance on DSP blocks.

One inherent limitation of MSDF arithmetic is the presence of initial delay ($\delta$), a fixed number of cycles that must elapse before the first digit of the result becomes available, increasing the computation latency. Traditionally, the MSDF inner product unit is developed using a multiplier followed by an adder tree structure to accumulate partial products \cite{shafique2024cnn}. In such implementations, each stage of the adder tree introduces its own delay, compounding the overall latency. To mitigate this issue, we propose to use a MSDF inner product computation with merged multiplication and addition (MMA) unit \cite{gorgin2023efficient}. This architecture eliminates the need for a separate adder tree and thus removes the associated initial delays, resulting in a more efficient datapath with reduced latency in MSDF-based computation systems.

\begin{figure}[htbp]
\centerline{\includegraphics[width=0.8\linewidth]{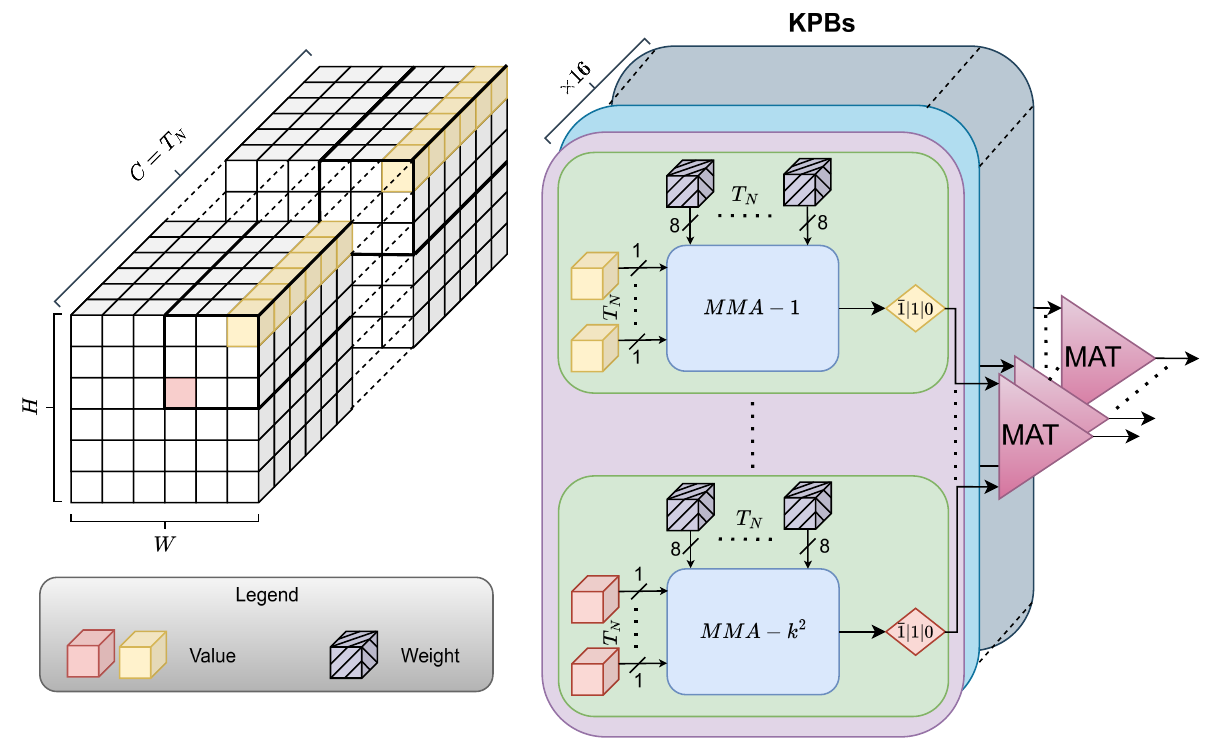}}
\caption{System-level architecture of the proposed U-Net convolution accelerator. Each merged multiplication-addition (MMA) unit computes the inner product across $T_N$ input channels using bit-serial activations and parallel $n$-bit weights. A group of $k^2$ MMAs forms a Kernel Processing Block (KPB), handling the $k\times k$ convolution over the spatial window. The outputs of the $k^2$ MMA units are summed using an MSDF adder tree to produce a single partial sum corresponding to one output pixel. The complete system instantiates 16 parallel KPBs, enabling the computation of 16 output pixels per clock group.}
\label{fig:system}
\end{figure}

\section{Proposed Design}
\subsection{Overall Architecture}
The overall architecture of the proposed design is shown in Fig.~\ref{fig:system}. The core computation unit of the accelerator is the MSDF arithmetic-based merged multiplication and addition unit (MMA), which computes the inner product across $T_N=32$ input channels. The detailed working principle of MMA is explained in Section~\ref{sec:PE}. To implement the $k\times k$ ($k=3$) kernel convolutions of U-Net efficiently, $k\times k = 9$ MMA units are grouped to form a column. Each MMA unit corresponds to one spatial location within the $k\times k$ convolution window but processes the same $N$ input channels in parallel. Outputs from the nine MMA units are combined using a $\lceil \log_2(k^2)\rceil$-stage MSDF adder tree (MAT) to form a Kernel Processing Block (KPB), preserving the digit-serial nature for continuous streaming and overlapping computation with communication. Collectively, the KPB computes the convolution over a $k\times k\times T_N$ volume, such that:
\begin{equation}
    \sum_{j=1}^{k\times k} S_j = S_{\text{KPB}} =  \sum_{j=1}^{k\times k} \sum_{i=1}^{T_N} \sum_{b=0}^{n} a_{i,j}^{(b)} \times w_{i,j}.
\end{equation}
To improve throughput, the architecture instantiates $16$ parallel KPBs, each working on independent regions of the input feature map. This allows the system to compute $16$ output pixels in parallel, improving performance and enabling real-time processing.

At the system level, multiple Kernel Processing Blocks can be instantiated to further improve throughput and efficiently utilize FPGA resources. This modular design is naturally suited to the U-Net architecture, which commonly features convolutional layers with $3\times 3$ kernels and input and output channels in multiples of $32$. However, it is general enough to accommodate other CNN architectures, as $3\times 3$ kernels and multiples of $32$ channels are very common. In special cases, larger kernels can be decomposed or the FPGA can be reconfigured. The full system includes pipelining, initial delays, and control overhead, leading to a total latency of $26$ cycles per output from a MMA.

\subsection{MSDF-based Merged Multiplication and Addition Unit}\label{sec:PE}
Recently, researchers have utilized MSDF arithmetic-based multipliers and adders to develop inner product computation units to accelerate convolution computation \cite{ibrahim2024echo,shafique2024cnn}. The properties of MSDF arithmetic have resulted in significant improvements in performance compared to conventional arithmetic-based designs for convolution computation. However, MSDF arithmetic-based computation units suffer from an inherent initial delay. This is a fixed delay of around $2$ to $5$ clock cycles, depending on the arithmetic function, and it becomes the bottleneck and adds to latency. For instance, to perform a convolution with a spatial depth of $N$ input channels, it takes $N$ instances of multipliers and $N-1$ instances of adders. The adders are arranged as a tree with $\lceil \log_2(N)\rceil$ levels. The initial delays of multiplier and adder are $\delta_{\times}$ and $\delta_{+}$ respectively, suggesting the algorithm requires $\delta_{\times}$ cycles to generate the first output from the multipliers and later each stage of the adder takes $\delta_{+}$ computation cycles to generate the output. The relation for the number of cycles to compute an $n$-bit input inner product is given as $\delta_{\times} + (\delta_{+} \times \lceil \log_2(T_N)\rceil) + p_{out}$, where $\delta_{\times}$ is the initial delay of MSDF multiplier, $\delta_{+}$ is the initial delay of adder, $T_N$ is the number of input channels being processed in parallel, and $p_{out}$ is the output precision, which for the $n=8$-bit inner product is: $p_{out} = (2\times n)+\lceil \log_2(T_N)\rceil = 16+5=21$.

\begin{figure}[htbp]
    \centerline{\includegraphics[width=0.75\linewidth]{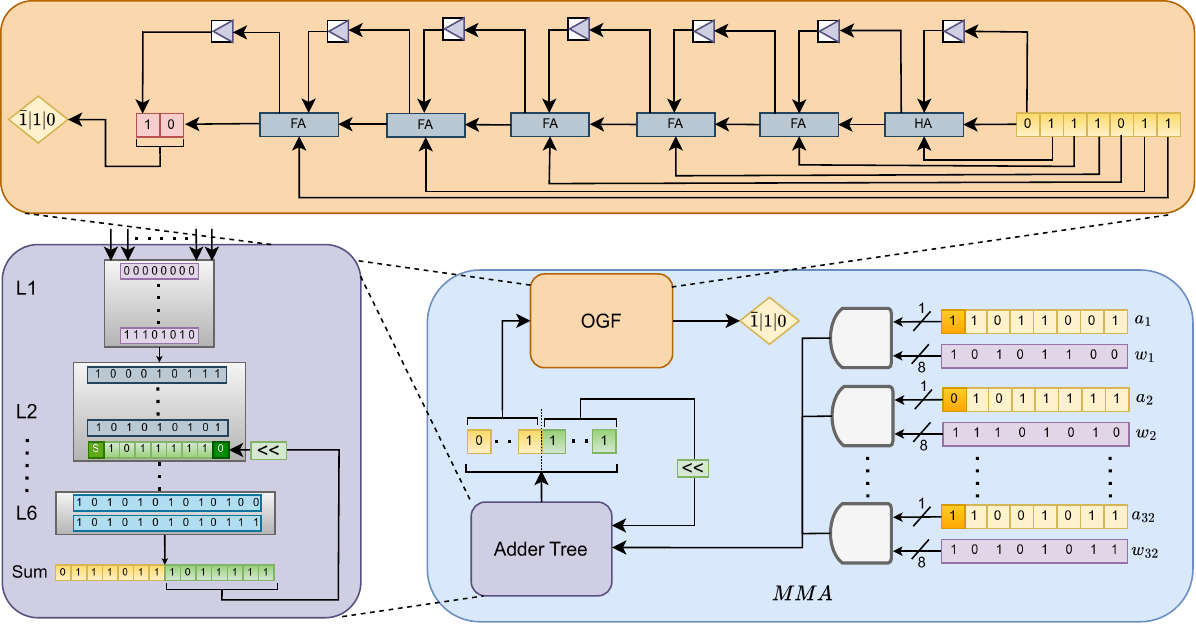}}
    \caption{Merged Multiplication Addition Architecture. The MMA unit consists of a 32-element AND gate array designed to process 8-bit quantized activations (serially streamed) and weights (available in parallel). The partial products generated by the array are forwarded to an adder tree that performs parallel accumulation of these products along with a residual carried over from the previous computation cycle. The most significant half of the accumulated result is passed to the output generation block OGF, which comprises a chain of 2-input full adders responsible for producing the next output digit in a digit-serial manner.}
\label{fig:MMA}
\end{figure}

In the proposed design, the multiplication and addition is fused to form a merged multiplication-addition unit, as shown in Fig.~\ref{fig:MMA}. The multiplication is implemented using bitwise AND gates: if the $b^\text{th}$ bit of the activation $a_i^{(b)} = 1$, the corresponding 8-bit weight $w_i$ is selected; otherwise, the output is zero. At each clock cycle, the MMA receives a 32-bit activation vector $\mathbf{a}^{(b)} = [a_1^{(b)}, a_2^{(b)}, \ldots, a_{32}^{(b)}]$, representing the $b^\text{th}$ bit (starting from the most significant bit) of activations from 32 channels, along with a parallel 8-bit weight vector $\mathbf{w} = [w_1, w_2, \ldots, w_{32}]$.
Element-wise bitwise multiplication is performed between $\mathbf{a}^{(b)}$ and $\mathbf{w}$, and the resulting weights are summed to generate a partial sum for the current bit plane. Over 8 cycles (corresponding to 8 activation bits), the MMA accumulates these partial sums to compute the full 8-bit by 8-bit multiplication result. Mathematically, the final output is:
\[
S = \sum_{b=0}^{7} \sum_{i=1}^{32} a_i^{(b)} \times w_i.
\]

\renewcommand{\arraystretch}{1.1}
\begin{table}[!ht]
\caption{Comparative analysis of multiple platforms across key performance indicators}
\begin{center}
\begin{tabular}{lllllll}
\hline \hline
Design & Bit-Parallel \cite{zhang2015optimizing} & Bit-Serial \cite{lee2018unpu} & MSDF \cite{ibrahim2024echo} & GPU & CPU & Proposed \\ \hline
Frequency (MHz) & 100 & 100 & 100 & ------- & 2200 & 100 \\ \hline
Time (ms) & 57.20 & 232.26 & 133.94 & 7.31 & 58.42 & 53.25 \\ \hline
Throughput (GOPS) & 49.30 & 12.14 & 21.05 & 385.99 & 48.27 & 52.95 \\ \hline
GOPS/W & 2.65 & 0.88 & 3.01 & 5.51 & 1.93 & 15.14 \\ \hline
Area Efficiency GOPS/Slice (10$^{-4}$) & 10.59 & 3.98 & 2.61 & ------- & ------- & 17.43 \\ \hline
Energy Efficiency (mJ) & 1064.43 & 3210.81 & 1644.77 & 511.35 & 1460.48 & 186.20 \\ \hline \hline
\end{tabular}
\end{center}
\label{tab:results}
\end{table}

Since each multiplication yields a 16-bit result (8-bit activation $\times$ 8-bit weight), and summing across 32 channels, the final output has a bit growth to $(\lceil \log_2(32)\rceil + 16) = 5 + 16 = 21$ bits to avoid overflow. The partial products generated in each cycle from the AND gates are accumulated using a carry-propagate adder tree, which, when synthesized, efficiently leverages FPGA carry-chain resources to minimize addition delay and maximize performance. The latency of the proposed method in terms of clock cycles is given as in relation \eqref{eq:CC}.
\begin{equation} \label{eq:CC}
    \Big(\delta_{\times+}+p_{out}+\big\lceil \log_2(T_N)\big\rceil\Big)\times
    \Big\lceil\frac{No.\;of\;Conv}{KPBs}\Big\rceil \times
    \Big\lceil \frac{N}{T_N}\Big\rceil,
\end{equation}
where $\delta_{\times+}$ corresponds to the initial delay of MMA unit and is equal to $2$, $p_{out}$ is the output precision bit width and is $(2\times n)+ \big\lceil \log_2(T_N)\big\rceil$, $KPBs$ are the kernel processing buffers, which are $16$ in this study, $N$ is the number of input channels in a certain layer, and $T_N$ is the input tiling factor which is $32$. Relation \eqref{eq:NConv} gives the number of convolutions in a layer.
\begin{equation}\label{eq:NConv}
\left\lfloor \left( \frac{R + 2P - k}{S} \right) + 1 \right\rfloor \times
\left\lfloor \left( \frac{C + 2P - k}{S} \right) + 1 \right\rfloor \times
\left\lceil \frac{M}{T_M} \right\rceil,
\end{equation}
where $R$ and $C$ is the height and width of the output feature map, $S$ is the stride, $P$ corresponds to padding, $M$ and $T_M$ corresponds to the number of output channels and tiling factor of the output channels respectively. In this study $T_M=1$.

In order to generate the output of MSDF based on only partial information about the input bits, which are gathered during initial cycles without generating output (initial delay $\delta$), the algorithm requires flexibility in the digit set. This flexibility allows the algorithm to correct the overall result in subsequent cycles if a wrong output is produced in the current cycle. To achieve this flexibility, the redundant number system (RDNS) is employed, which allows multiple representations of a numerical value. In this work, we employ signed digit (SD) redundant number system of radix-2 with digits drawn from the set $\{\overline{1},0,1\}$. Each digit is encoded using two bits: $\overset{+}{x}$ and $\overline{x}$, where $\overset{+}{x}$ indicates the positive component and $\overline{x}$ the negative component of the digit. The digit $0$ is represented by the bit patterns $01$ or $10$, digit $1$ is encoded as $11$, and digit $\overline{1}$ is encoded as $00$. It is noteworthy that we employ Inverted Encoding of Negabit (IEN) in this study, where a logical $0$ corresponds to the arithmetic value $-1$ and a logical $1$ to $+1$ \cite{jaberipur2008constant}. This encoding is the inverse of the conventional two's complement representation, where the most significant bit is $0$ for positive values and $1$ for negative ones. This reversed encoding convention simplifies sign handling and is beneficial in the context of left-to-right and digit-serial accumulation.

As shown in Fig.~\ref{fig:MMA}, $T_N=32$ partial products from AND gates are added in the adder tree in each cycle, requiring $\lceil \log_2(T_N)\rceil$ adder stages. However, the algorithm also demands the inclusion of a partial result from the previous cycle, referred to as the residual, in the addition process. Consequently, the adder tree must accommodate $T_N+1$ inputs, increasing the number of required adder stages to $\lceil \log_2(T_N+1)\rceil$. The adder produces a result of $v = n + \lceil \log_2(32+1)\rceil = 14$ bits in each cycle. To generate the output, the $t=\lceil \log_2(32+1)\rceil + 1 = 7$ most significant bits of this partial sum are forwarded to the output digit generation function (OGF). The remaining $r = v - t = 7$ least significant bits are treated as the residual and fed back into the second stage of the adder tree in the next cycle. Before feedback, the residual is left-shifted by one bit to align with the weight of the next incoming partial products. Additionally, its least significant bit is forced to $0$, and sign extension is applied to fill the most significant digit positions, ensuring correct accumulation across cycles.

\section{Implementation and Results}
The U-Net model was initially implemented in PyTorch 2.5.0 and subsequently quantized using the FBGEMM \cite{fbgemm} backend to enable efficient fixed-point inference suitable for hardware acceleration. The accelerator was implemented on a Xilinx Zynq-7020 SoC using Verilog RTL.
The proposed accelerator has been compared with the state-of-the-art conventional bit-parallel \cite{zhang2015optimizing}, bit-serial \cite{lee2018unpu}, contemporary MSDF-based accelerator \cite{ibrahim2024echo}, GPU, and CPU implementations in terms of throughput, energy efficiency, and area utilization (Tab.~\ref{tab:results}). Operating at $100$ MHz, our design achieves $1.07\times$ higher throughput than the conventional bit-parallel and $4.36\times$ higher than the bit-serial design. Compared to the MSDF-based CNN accelerator, it delivers $2.52\times$ greater throughput. Although the GPU achieves the highest absolute throughput, the proposed design is more energy efficient, with $2.7\times$ better than the GPU and over $7\times$ better than the CPU, making it highly suitable for power-constrained environments. The area efficiency is also improved, achieving $1.65\times$ higher GOPS per slice compared to conventional bit-parallel designs and nearly $6.7\times$ higher than MSDF-based accelerator. Furthermore, the energy consumption per operation batch is significantly reduced, being $3.8\times$ lower than the GPU and $10.8\times$ lower than the CPU, making the design ideal for embedded and edge applications where energy constraints are critical.

\section{Conclusion and Future Work}
In this work, we presented a novel convolution accelerator that improves throughput, energy efficiency, and area utilization over conventional bit-parallel, bit-serial, MSDF-based designs, as well as GPU and CPU implementations. By leveraging parallelism and efficient arithmetic, the design delivers high performance at moderate frequency with significant power savings, making it well-suited for energy-constrained edge applications. In future work, we plan to add early termination to reduce computation time and energy by skipping unnecessary operations, and to scale the architecture with more tiles to support greater parallelism and meet the demands of more complex CNNs.

\bibliographystyle{IEEEtran}
\bibliography{Reference}

\end{document}